%% file: computing.tex
\documentclass[10pt,conference]{IEEEtran} 
\IEEEoverridecommandlockouts 
\usepackage{tikz}
\tikzstyle{arrow} = [thick,->,>=stealth]
\usetikzlibrary{decorations.pathreplacing}
\usepackage{amsmath,amsthm,amssymb,amsfonts,dsfont,bm}

\usepackage[bookmarksopen=true,pagebackref=true]{hyperref}
\usepackage{cleveref}
\usepackage{caption}
\usepackage[labelformat=simple]{subcaption}

\usepackage[noadjust]{cite}
\allowdisplaybreaks[2]
\usepackage{stfloats}
\usepackage{enumerate}

\theoremstyle{plain}
\newtheorem{theorem}{Theorem}

\newtheorem{lemma}{Lemma}

\theoremstyle{definition}
\newtheorem{definition}{Definition}
\newtheorem{example}{Example}

\theoremstyle{remark}
\newtheorem{remark}{Remark}

\begin{document}

\title{Lossy Computing with Side Information via  Multi-Hypergraphs}


\author{
Deheng Yuan, Tao Guo, Bo Bai, and Wei Han
}

\maketitle

%

\begin{abstract}
We consider a problem of coding for computing, where the decoder wishes to estimate a function of its local message and the source message at the encoder within a given distortion.
We show that the rate-distortion function can be characterized through a characteristic multi-hypergraph, 
which 
simplifies the evaluation of the rate-distortion function.

\end{abstract}


\input{A0}

\input{A1}

\input{A2}

\input{A3}
\input{A4}
\input{A5}

\bibliographystyle{IEEEtran}
\bibliography{computing}

%

\end{document}

%% file: A0.tex
\section{Introduction}
\label{sec:introduction}

Consider the lossy computing problem with side information.
Let $f$ be a function of two sources $X$ and $Y$, which are observed by the encoder and the decoder respectively. 
Upon receiving a message from the encoder, the decoder makes an estimate of the function $f(X,Y)$. 
Our goal is to determine the minimum number of transmitted  bits so that the estimation is within a given distortion.

Wyner and Ziv studied the case of $f(x,y)=x$ in~\cite{WynerZiv1976}, which is known as the rate-distortion problem with side information.
The rate-distortion function for a general $f$ was given by Yamamoto~\cite{Yamamoto1982} in terms of an auxiliary random variable, 
for which however the intuitive meaning is not clear.

The notions of {\it graph entropy} and {\it characteristic graph} were introduced by K\"{o}rner~\cite{Korner1973} and Witsenhausen~\cite{Witsenhausen1976} for zero-error coding problems.
Orlitsky and Roche~\cite{Orlitsky2001} extended the tools and obtained a graph-based characterization of the minimum rate for lossless computing with side information. 
The auxiliary random variable involved therein is clearly represented by the independent set of a characteristic graph.

To better understand the lossy computing problem, a natural generalization of the graph entropy approach in~\cite{Orlitsky2001} was given in ~\cite{Doshi2007} and \cite{Doshi2010} by defining the $D$-characteristic graph, 
where an efficient but suboptimal coding scheme was obtained.
In \cite{Basu2020} and \cite{Basu2022}, Basu, Seo and Varshney 
generalized the independent sets to hyperedges and defined an $\epsilon$-characteristic hypergraph, 
where a hyperedge exists only when the corresponding source values induce a distortion on $f$ less than or equal to~$\epsilon$. 
The rate-distortion function was characterized for a limited class of distortion measure whose average represents the probability that the distance between $f$ and the reconstruction is larger than a given distortion level. 
Their generalization, however, cannot cope with general distortion measures.

In the current paper, we further generalize the characteristic hypergraph to characteristic multi-hypergraph by allowing a larger set of hyperedges that is independent of the distortion. 
The rate-distortion function can be characterized for any general distortion measures, wherein the auxiliary random variable can be constructed from the hyperedges in the characteristic multi-hypergraph. 
The proposed multi-hypergraph also provide a graph-based optimal coding scheme. 
Our result naturally subsumes that in~\cite{Basu2020} as a special case by specifying a distortion measure. 

In \Cref{PS} we formulate the problem and describe some preliminaries. 
The main results are given in~\Cref{sec:results}. 
In \Cref{sec:coding} we show the advantage of the probabilistic multi-hypergraph in evaluating the rate-distortion function and designing optimal coding schemes.
Essential proofs can be found in \Cref{proof} and we conclude the paper in \Cref{sec:conclusion}.

%% file: A1.tex
\section{Problem Formulation and Preliminaries}\label{PS}
\subsection{Problem Formulation}
Denote a discrete random variable by a capital letter and its finite alphabet by the corresponding calligraphic letter, e.g., $X\in\mathcal{X}$ and $\hat{Z}\in\hat{\mathcal{Z}}$.
We use the superscript $n$ to denote an $n$-sequence, e.g., $X^n=(X_i)_{i = 1}^n$.
Let $(X_i,Y_i) \sim p(x,y),i\in\{1,2,\cdots,n\}$ be i.i.d. random variables distributed over $\mathcal{X}\times\mathcal{Y}$. 
Without loss of generality, assume $p(x) > 0$, $\forall x \in \mathcal{X}$ throughout this paper.

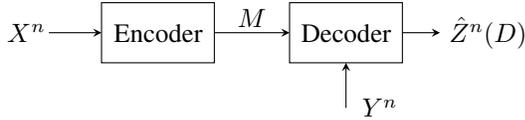
\begin{figure}
\centering
\begin{tikzpicture}
	\node at (0.5,0) {$X^n$};
	\draw[->,>=stealth] (0.8,0)--(1.5,0);
	
	\node at (5.2,-1.0) {$Y^n$};
	\draw[->,>=stealth] (4.75,-1.0)--(4.75,-0.4);
	
	\node at (2.25,0) {Encoder};
	\draw (1.5,-0.4) rectangle (3.0,0.4);
	
	\draw[->,>=stealth] (3.0,0)--(4,0);
	\node at (3.5,0.2) {$M$};
	
	\node at (4.75,0) {Decoder};
	\draw (4.0,-0.4) rectangle (5.5,0.4);
	
	\draw[->,>=stealth] (5.5,0)--(6.0,0);
	\node [right] at (6.0,0) {$\hat{Z}^n (D)$};
\end{tikzpicture}
\caption{Lossy computing with side information}
\label{fig:system_model}
\vspace{-0.2cm}
\end{figure}

Consider the lossy computing problem with decoder side information depicted in Fig.~\ref{fig:system_model}. 
The source messages $X^n$ and $Y^n$ are observed by the encoder and the decoder, respectively. 
Let $f : \mathcal{X} \times \mathcal{Y} \to \mathcal{Z}$ be the function to be computed and $d: \mathcal{Z} \times \hat{\mathcal{Z}} \to [0,\infty)$ be a distortion measure. 
Denote $f(X_i,Y_i)$ by $Z_i$ for $1\leq i\leq n$. 
Without ambiguity, we abuse the notation of $f$ and $g$ to denote their vector extensions, and define 
\begin{align*}
f(x^n,y^n) &= \big(f(x_i,y_i)\big)_{i = 1}^n, \\
d(z^n,\hat{z}^n) &= \frac{1}{n}\sum_{i = 1}^n d(z_i,\hat{z}_i).
\end{align*}

An $(n,2^{nR})$ code is defined by an encoding function $$g_e: \mathcal{X}^n \to \{1,2,...,2^{nR}\}$$ 
and a decoding function 
$$g_d:\{1,2,...,2^{nR}\} \times \mathcal{Y}^n \to \hat{\mathcal{Z}}^n.$$
Then the decoded messages are $\hat{Z}^n = g_d(g_e(X^n),Y^n)$.

A rate-distortion pair $(R,D)$ is said {\it achievable} if there exists an $(n,2^{nR})$ code such that 
\[
\varlimsup_{n \to \infty} \mathbb{E}[d({Z}^n,\hat{Z}^n)] \leq D.
\]
We define the rate-distortion function $R(D)$ to be the infimum of all the achievable rates such that $(R,D)$ is achievable.

\subsection{Existing Results}

Yamamoto obtained the rate-distortion function in \cite{Yamamoto1982}. 
We characterize the result in the following lemma. 

\begin{lemma}\label{lem1}
The rate-distortion function is given by 
\begin{equation}\label{basic}
R(D) = \min_{\substack{U-X-Y \\ \exists g: \mathbb{E}[d(f(X,Y),g(U,Y))] \leq D}}I(X;U|Y). 
\end{equation}
Note that the minimum is taken over all random variables $U$ such that $U-X-Y$ forms a Markov chain, and there exists a decoding function $g : \mathcal{U} \times \mathcal{Y} \to \hat{\mathcal{Z}}$ such that 
\[
\mathbb{E}[d(f(X,Y),g(U,Y))] \leq D.
\]
Moreover, the size of the alphabet can be bounded by 
\begin{equation}\label{cardinality bound}
|\mathcal{U}| \leq |\mathcal{X}| +1.
\end{equation}
\end{lemma}
It was further noted in \cite{Yamamoto1982} that the above rate-distortion function is decreasing and convex in $D$ for $D\in[0,\infty)$.


\subsection{Characteristic Multi-Hypergraph}
A {\it multi-hypergraph}~\cite{Bretto2013} consists of a pair $G_m = (V,E)$, where $V$ is a finite vertex set and $E$ is a family of subsets of~$V$ that allows multi-hyperedge, i.e., the same subset of $V$ may appear more than once in $E$. 
If there is no multi-hyperedge, i.e., any subset of $V$ can appear at most once in $E$, then we call $G_m$ a {\it hypergraph}.\footnote{A hypergraph is always a multi-hypergraph. Note that the multi-hypergraph defined here is simply called a hypergraph in some literature, e.g.,~\cite{Bretto2013}.}
In other words, the edge set $E$ of a hypergraph is a subset of the power set of $V$.

Without loss of generality, we consider only the multi-hypergraphs satisfying 
\[
V \neq \emptyset, E \neq \emptyset, \text{ and }w \neq \emptyset \text{ for all } w \in E.
\]
For simplicity, let $\mathcal{P}(V) = \{w| w \subseteq V, w  \neq \emptyset \}$.

For any hypergraph $G=(V,E)$ and a finite set $J$, we can regard $E \times J$ as the set of hyperedges of a multi-hypergraph $G_m=(V,E \times J)$ by duplicating $|J|$ times of each hyperedge $w\in E$ as $w_{j_1},w_{j_2},\cdots,w_{j_{|J|}}$.

Let $\hat{z}_y\in\hat{\mathcal{Z}}$ denote the realization of $\hat{Z}_{Y}$ corresponding to the realization of $Y=y\in\mathcal{Y}$. 
Then 
\begin{equation}
    \hat{z}_{\mathcal{Y}} \triangleq (\hat{z}_{y_1}, \hat{z}_{y_2},\cdots, \hat{z}_{y_{_{|\mathcal{Y}|}}})
\end{equation}
can be viewed as a mapping from $\mathcal{Y}$ to $\hat{\mathcal{Z}}$, and $\hat{Z}_{\mathcal{Y}}$ is the corresponding random variable. 
We define $$\hat{\mathcal{Z}}^{\mathcal{Y}}= \{\hat{z}_{\mathcal{Y}}: \hat{z}_{y_i} \in \hat{\mathcal{Z}} \text{ for }i=1,2,\cdots,|\mathcal{Y}|\}$$ to be the collection of all such mappings, or equivalently, the collection of all tuples with each component $\hat{z}_{y_i}$ taking values from $\hat{\mathcal{Z}}$.

In light of a hypergraph $G=(\mathcal{X},\mathcal{P}(\mathcal{X}))$, we can define the {\it characteristic multi-hypergraph}\footnote{Note that there is a simple transition from multi-hypergraphs to bipartite graphs, so our results can also be interpreted in terms of bipartite graphs. We omit the details here and will discuss in future work. } as follows.

\begin{definition}
The {\it characteristic multi-hypergraph} for the lossy computing problem is defined by
\[
G_{m,\chi}=(\mathcal{X},\mathcal{P}(\mathcal{X}) \times \hat{\mathcal{Z}}^{\mathcal{Y}}).
\]
\end{definition}

For a random variable (hyperedge) $W \in \mathcal{P}(\mathcal{X})$, we say $X \in W$ if $\mathbb{P}[X \in W] = 1$. In other words, $p(w,x)>0$ only if $x \in w$. 

Note that for zero distortion case, 
$G_{m,\chi}$ reduces to a characteristic hypergraph $G^*=(\mathcal{X},\Gamma_d)$,  where the hyperedge set $\Gamma_d$ will be defined in \Cref{SC}.

%% file: A2.tex
\section{Main Results}
\label{sec:results}

\subsection{General Results}\label{GR}
We first establish the main result that fully solves the lossy computing problem with general distortion measures. 
Some interesting reductions will be discussed subsequently.
The proofs of our results can be found in~\Cref{proof}.

\begin{theorem}\label{Thm1}
For any $D \geq 0$, the rate-distortion function $R(D)$ is given by 
\begin{equation}\label{eqthm1}
R(D) = \min I(X;\tilde{W}|Y),
\end{equation}
where the minimum is taken over all the random variables $\tilde{W} = (W,\hat{Z}_{\mathcal{Y}})$ satisfying $\tilde{W}-X-Y$, $X \in W \in \mathcal{P}(\mathcal{X})$, and $\mathbb{E}[d(f(X,Y),\hat{Z}_Y)] \leq D$.
\end{theorem}

\begin{remark}\label{remark:cardinality}
From the proof in \Cref{proof} (c.f. \eqref{Thm1cardinality}), we see that there always exists some $\tilde{W}$ which achieves the minimum in \Cref{Thm1} and has a sparse distribution (a small support).
More precisely, there are at most $|\mathcal{X}|+1$ of $\tilde{w} \in \tilde{\mathcal{W}}$
satisfying $p(\tilde{w})>0$, 
which shows that the support of $\tilde{W}$ is relatively small despite the large cardinality of $\tilde{\mathcal{W}}$.
The sparsity is inherited by the auxiliary random variables in \Cref{Thm2,Thm3}, which can be seen from their proofs.
\end{remark}

\begin{remark}
Note that $W\in\mathcal{P}(\mathcal{X})$ is a hyperedge in the hypergraph $G=(\mathcal{X},\mathcal{P}(\mathcal{X}))$ and $\tilde{W}\in \tilde{\mathcal{W}} = \mathcal{P}(\mathcal{X}) \times \hat{\mathcal{Z}}^{\mathcal{Y}}$ is a hyperedge in the characteristic multi-hypergraph $G_{m,\chi}$.  
\end{remark}

In addition to solving the rate-distortion function, the characteristic multi-hypergraph also induces a coding protocol. 
We explain the brief ideas as follows, and the details are illustrated through an example in \Cref{sec:coding}.
Each hyperedge $\tilde{w}$ is composed of two parts, $w \subseteq \mathcal{X}$ and the candidate recoveries $\hat{z}_{\mathcal{Y}}$.
While observing $x \in \mathcal{X}$, the encoder encodes it into $\tilde{w}=(w,\hat{z}_{\mathcal{Y}})$ such that $x \in w$, with probability $p(\tilde{w}|x)$. 
Upon receiving~$\tilde{w}$ and observing $y$, the decoder generates an estimate $\hat{z}_y$ by looking up the candidate recovery part of $\tilde{w}$ to find the component with index $y$.


\subsection{Direct Reduction}\label{DR}
We see from \Cref{remark:cardinality} that the support of $\tilde{W}$ can be relatively small. 
Now we simplify \Cref{Thm1} by specifying a more concise auxiliary random variable in the following theorem. 
\begin{theorem}\label{Thm2}
For any $D \geq 0$, the rate-distortion function $R(D)$ can be characterized by 
\begin{equation}
R(D) = \min I(X;\tilde{W}|Y),
\end{equation}
where the minimum is taken over all random variables $\tilde{W} = \hat{Z}_{\mathcal{Y}}$ 
satisfying $\tilde{W}-X-Y$ and  $\mathbb{E}[d(f(X,Y),\hat{Z}_Y)] \leq D$.
\end{theorem}

\begin{remark}
Consider the special case that $Y$ is a constant and $f(x) = x$, then \Cref{Thm2} reduces to Shannon's rate-distortion theorem.
\end{remark}

\begin{remark}
\Cref{Thm2} implies that the candidate recovery part~$\hat{Z}_{\mathcal{Y}}$ contains enough information and suffices to recover the subset part $W$ of the hyperedge in the multi-hypergraph, 
which is also shown in the proof of \Cref{Thm2}.
\end{remark}

With the result in \Cref{Thm2}, one may argue the significance of multi-hypergraphs and the characterization in \Cref{Thm1}.
However, we will see in the following section that the subset part $W$ in the hyperedge plays an important role in the special case of $D=0$, where the candidate recovery part $\hat{Z}_{\mathcal{Y}}$ can even be deleted.
Moreover, the explicit meaning of $W$ in the hyperedge can help to solve the optimization problem, which will be demonstrated through an example in \Cref{sec:coding}.

\subsection{Zero Distortion Case}\label{SC}
Consider the case of $D=0$, and assume that  for each $z \in \mathcal{Z}$, there exists some $\hat{z} \in \hat{\mathcal{Z}}$ such that $d(z,\hat{z}) = 0$.
Then the support of the auxiliary random variable $\tilde{W} = (W,\hat{Z}_{\mathcal{Y}})$ can be determined by only local properties of the hyperedges,
which are described by the following definition.

\begin{definition}
For each $w \subseteq \mathcal{X}$
and $y\in\mathcal{Y}$, let $w_y = \{f(x,y): x \in w \text{ and } p(x,y)>0 \}$.
Then $\Gamma_d$ is the collection of all $w$ satisfying the following conditions: 
\begin{enumerate}[(i)]\label{properties}
\item $w \neq \emptyset $, in other words,
$w \in \mathcal{P}(\mathcal{X})$.

\item For each $y \in \mathcal{Y}$, there exists some $\hat{z} \in \hat{\mathcal{Z}}$, such that $w_y \subseteq B(\hat{z},0)$, where $B(\hat{z},\delta)\triangleq \{z \in \mathcal{Z}: d(z,\hat{z})  \leq \delta\}$ for $\delta \geq 0$.
\end{enumerate}
\end{definition}

We see from above that each $x \in \mathcal{X}$ must be contained in some $S \in \Gamma_d$ since $x \in \{x\} \in \Gamma_d$. 
With the definition of $\Gamma_d$, \Cref{Thm1} reduces as follows, where the characteristic multi-hypergraph $G_{m,\chi}$ reduces to a hypergraph $G^*=(\mathcal{X},\Gamma_d)$ with much less hyperedges.

\begin{theorem}\label{Thm3}
For $D=0$, the rate-distortion function is 
\begin{equation}
R(0) = \min I(X;W|Y),
\end{equation}
where the minimum is taken over all random variables $W$ 
satisfying $W-X-Y$ and $X \in W \in \Gamma_d$.
\end{theorem}

\begin{remark}
Let $d_{\epsilon}(z,\hat{z}) = \mathds{1}\{ d(z,\hat{z}) > \epsilon \}$ for any $\epsilon \geq 0$, where $\mathds{1}$ denotes the indicator function. 
Then the main result of~\cite[Theorem 3]{Basu2020} can be obtained by applying \Cref{Thm3} to the distortion measure $d_{\epsilon}$.
\end{remark}

\begin{remark}
Assume $\mathcal{Z} = \hat{\mathcal{Z}}$ and $d$ satisfies 
\begin{equation}
    d(z,\hat{z})= 0 \iff z=\hat{z}.
\end{equation}
Then the hyperedges in $\Gamma_d$ become independent sets of the characteristic graph in \cite{Orlitsky2001} and \Cref{Thm3} reduces to the significant results of Theorem~2 therein.
\end{remark}

\Cref{Thm3} is a reduction of \Cref{Thm1}, in the following, we discuss the main differences between them, which may 
help explain why the zero distortion case in \Cref{Thm3} and \cite{Orlitsky2001}\cite{Basu2020}\cite{Basu2022} is much simpler and 
give further insights on why multi-hypergraphs are essential to fully solve the lossy computing problem. 

Firstly, the feasible region of hyperedges in \Cref{Thm3} is limited to $\Gamma_d$ and is much smaller than that in \Cref{Thm1}, which is the whole hyperedge set. 
The intuition behind is the ``zero effect" for $D=0$, i.e., for each hyperedge $\tilde{w}$ with $p(\tilde{w})>0$, the distortion induced by $\tilde{w}$ must be zero.
However, for $D>0$, even hyperedges inducing a distortion larger than~$D$ are still possible, since the average distortion is of final concern.


{
Secondly, in the general problem considered in \Cref{Thm1}, fix a $y\in\mathcal{Y}$, for any hyperedge $w\subseteq \mathcal{X}$, the induced reconstruction is required to take different values in $\hat{\mathcal{Z}}$ in order to achieve a smaller average distortion. 
To illustrate the correspondence between the hyperedges and the reconstruction, }
we need a hyperedge in the original hypergraph $(\mathcal{X},\mathcal{P}(\mathcal{X}))$ to repeat multiple times which are distinguished by their different candidate recovery in the characteristic multi-hypergraph $G_{m,\chi}$,
and the candidate recovery $\hat{Z}_{\mathcal{Y}}$ for each $w\in\mathcal{P}(\mathcal{X})$ can only be determined in the minimization process. 
However, in \Cref{Thm3}, the candidate recovery for each hyperedge $w \in \Gamma_d$ is simply chosen to be the $\hat{z}$ that induces zero distortion.



%% file: A3.tex
\subsection{Example}\label{sec:coding}
We use an example to illustrate how the optimization in the rate-distortion function can be simplified by the explicit meaning of hyperedges in the multi-hypergraph.

\begin{example}[Online card game]
Alice and Bob each randomly select one out of three cards labeled 1, 2, and 3 without replacement.
Alice agrees to help Bob determine who selected the card with a larger label.

Denote the label of Alice's card  by $X$, and Bob's by $Y$. 
Then  $(X,Y) \sim p(x,y)$ with $p(i,j) = \frac{1}{6}(1-\delta_{i,j}),i,j=1,2,3$,
and $f(x,y) = \mathds{1}\{x>y\}$, where $\delta_{i,j}=1$ if $i=j$ and 0 otherwise. 
Let $d$ be the Hamming distortion on $\mathcal{Z} = \hat{\mathcal{Z}}  = \{ 0,1 \}$. 
Then we compute the rate-distortion function as follows.

Assume $0 \leq D < \frac{1}{6}$, since it is easily seen that $R(D) = 0$ for $D \geq  \frac{1}{6}$. 
By the decreasing and convex properties of $R(D)$ for $D \geq 0$, we see $R(D)$ is strictly decreasing in $0 \leq D \leq \frac{1}{6}$.

In light of Theorem 2, we have 
\[
R(D)= \min_{\tilde{W}-X-Y, W = \hat{Z}_{\mathcal{Y}}, \mathbb{E}[d(f(X,Y),\hat{Z}_Y)] \leq D} I(X;\tilde{W}|Y).
\]

Consider the optimal $\tilde{W}$ that attains the rate-distortion function. 
Let $w(\tilde{w})= \{x \in \mathcal{X}: p(\tilde{w},x)>0 \}$.
Then for each $\tilde{w}$ such that $p(\tilde{w})>0$, $(w(\tilde{w}),\tilde{w})$ is a hyperedge in the characteristic multi-hypergraph $G_{m,\chi}$. 
For each $y$ with $p(\tilde{w},y)>0$, the recovery $\hat{z}_y$ must be optimal on $w(\tilde{w})$, otherwise replacing it with the optimal one can induce a smaller average distortion without increasing the mutual information, which contradicts the strictly decreasing property of $R(D)$. 
Similarly, by contradiction, the equality holds in the distortion constraint $\mathbb{E}[d(f(X,Y),\hat{Z}_Y)] \leq D$.

If $Y = 1$, then $X=2$ or $3$, and thus $X>Y$. So $\hat{Z}_1 = 1$, and similarly, $\hat{Z}_3 = 0$. For $\tilde{W}$ that achieves $R(D)$, only $\tilde{w}_0 = (1,0,0)$ and $\tilde{w}_1 = (1,1,0)$ can have positive probabilities.

For $i = 1,2,3$, let $p(\tilde{w}_0|x = i) = p_i$,
then $ p(\tilde{w}_1|x = i) =1-p_i$ and $\ 0 \leq p_i \leq 1$.
Since $Y$ and $\tilde{W}$ are mutually independent given $X$, we have 
\[
\begin{split}
&p(\tilde{w}_0,y = j,x=i) = \frac{1}{6} p_i (1-\delta_{ij}),
\\
&p(\tilde{w}_0,y = j,x=i) = \frac{1}{6} (1-p_i) (1-\delta_{ij}),\  i, j = 1,2,3 .
\end{split}
\]
The optimality condition on the subset $\{1,2,3\}$ for $Y = 2$ gives $p_1 \geq p_3$.
Moreover, the distortion constraint with equality implies 
$\frac{1}{6}(1-p_1+p_3) = D.$ 

The conditional mutual information is calculated as follows:
\begin{align*}
    &\hspace{-0.3cm}I(X;\tilde{W}|Y) = H(\tilde{W}|Y)-H(\tilde{W}|X)\\
    =&\frac{1}{3}(H(\frac{p_1+p_2}{2})+H(\frac{p_1+p_3}{2})+H(\frac{p_2+p_3}{2}))\\
    &-\frac{1}{3}(H(p_1)+H(p_2)+H(p_3))\\
    =& \frac{1}{3}[H(\frac{1-6D+p_2+p_3}{2})+H(p_3+\frac{1-6D}{2})\\
    &+H(\frac{p_2+p_3}{2}) -H(p_3+1-6D)-H(p_2)-H(p_3)]
\end{align*}

\begin{figure}
\centering
\begin{tikzpicture}
\node at (2,0.75) {$X$};
\node at (4,0.75) {$\tilde{W}$};
\node at (6.5,0.75) {$(\tilde{W},Y)$};
\node at (9,0.75) {$\hat{Z}$};
\node at (3,0.2) {$p(\tilde{w}|x)$};

\node at (2,0) {$1$};
\draw (2,0) circle(0.25);

\node at (2,-1) {$2$};
\draw (2,-1) circle(0.25);

\node at (2,-2) {$3$};
\draw (2,-2) circle(0.25);

\node at (4,-0.5) {$\tilde{w}_0$};
\draw (3.75,-0.75) rectangle (4.25,-0.25);

\node at (4,-1.5) {$\tilde{w}_1$};
\draw (3.75,-1.75) rectangle (4.25,-1.25);

\draw [->,>=stealth] (2.25,0)--(3.75,-0.5);
\draw [->,>=stealth] (2.25,0)--(3.75,-1.5);

\draw [->,>=stealth] (2.25,-1)--(3.75,-0.5);
\draw [->,>=stealth] (2.25,-1)--(3.75,-1.5);

\draw [->,>=stealth] (2.25,-2)--(3.75,-0.5);
\draw [->,>=stealth] (2.25,-2)--(3.75,-1.5);

\draw [->,>=stealth] (4.25,-0.5)--(5,-0.2);
\draw [->,>=stealth] (5,-0.2)--(5,0.1)--(5.7,0.1);
\node at (6.5,0.1) {$(\tilde{w}_0,y=1)$};
\draw [->,>=stealth] (5,-0.2)--(5.7,-0.2);
\node at (6.5,-0.2) {$(\tilde{w}_0,y=2)$};
\draw [->,>=stealth] (5,-0.2)--(5,-0.5)--(5.7,-0.5);
\node at (6.5,-0.5) {$(\tilde{w}_0,y=3)$};

\draw [->,>=stealth] (4.25,-1.5)--(5,-1.8);
\draw [->,>=stealth] (5,-1.8)--(5,-1.5)--(5.7,-1.5);
\node at (6.5,-1.5) {$(\tilde{w}_1,y=1)$};
\draw [->,>=stealth] (5,-1.8)--(5.7,-1.8);
\node at (6.5,-1.8) {$(\tilde{w}_1,y=2)$};
\draw [->,>=stealth] (5,-1.8)--(5,-2.1)--(5.7,-2.1);
\node at (6.5,-2.1) {$(\tilde{w}_1,y=3)$};

\node at (9,-0.5) {$1$};
\draw (9,-0.5) circle(0.25);

\node at (9,-1.5) {$0$};
\draw (9,-1.5) circle(0.25);

\draw [->,>=stealth] (7.4,0.1)--(8.75,-0.5);
\draw [->,>=stealth] (7.4,-0.2)--(8.75,-1.5);
\draw [->,>=stealth] (7.4,-0.5)--(8.75,-1.5);

\draw [->,>=stealth] (7.4,-1.5)--(8.75,-0.5);
\draw [->,>=stealth] (7.4,-1.8)--(8.75,-0.5);
\draw [->,>=stealth] (7.4,-2.1)--(8.75,-1.5);

\draw[decorate,decoration={brace,mirror}](2,-2.5) -- (3.56,-2.5);
\node at (3,-3) {probabilistic encoding};
\draw[decorate,decoration={brace,mirror}](4,-2.5) -- (9,-2.5);
\node at (7,-3) {recovery process};
\end{tikzpicture}
\caption{Coding scheme induced by multi-hypergraph}
\label{fig:example}
\vspace{-0.2cm}
\end{figure}
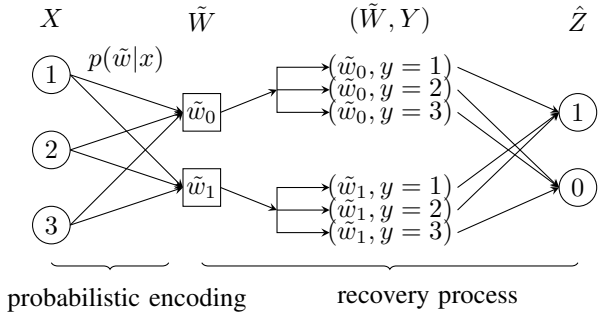

By the convexity, the minimum rate is obtained by differentiating the above function that $R(D)= \frac{2}{3}(H(\frac{1+6D}{4})-H(3D))$ for $0 \leq D < \frac{1}{6}$, and the minimum is achieved at $p_1 = 1-3D$, $p_2 = \frac{1}{2}$, $p_3 = 3D$.
The single-letter coding scheme is then obtained, and depicted in Fig.~\ref{fig:example}. Each single $X$ is encoded into one of the hyperedges $(\{1,2,3\},(1,0,0))$ and $(\{1,2,3\},(1,1,0))$ in the characteristic multi-hypergraph with the transition probability $p(\tilde{w}|x)$, and with $Y$ the decoder can recover the estimate  $\hat{Z}$. 


\end{example}

%% file: A4.tex
\section{Proofs}\label{proof}

\begin{IEEEproof}[Proof of \Cref{Thm1}]
 By \Cref{lem1}, we only need to show 
\[
\begin{split}
\min_{\substack{U-X-Y \\ \exists g: \mathbb{E}[d(f(X,Y),g(U,Y))] \leq D}}I(X;U|Y) = \min I(X;\tilde{W}|Y),
\end{split}
\]
where the right hand side is defined in \Cref{Thm1}.

We first prove ``$\leq$".
Suppose that $\tilde{W} = (W,\hat{Z}_{\mathcal{Y}})$
satisfies $\tilde{W}-X-Y$, $X \in W \in \mathcal{P}(\mathcal{X})$, and $\mathbb{E}[d(f(X,Y),\hat{Z}_Y)] \leq D$.
Let $U = \tilde{W}$, then $U-X-Y$ holds. We only need to find the function $g$, such that $\mathbb{E}[d(f(X,Y),g(\tilde{W},Y)] \leq D$.

Define $g(\tilde{w}, y) = \hat{z}_{y}$ for $\tilde{w} = (w,\hat{z}_{\mathcal{Y}}) \in \mathcal{P}(\mathcal{X}) \times \hat{\mathcal{Z}}^{\mathcal{Y}}$ and $y \in \mathcal{Y}$. 
By the assumption, we have
\[
\mathbb{E}[d(f(X,Y),g(\tilde{W},Y))] = \mathbb{E}[d(f(X,Y),\hat{Z}_Y)] \leq D.
\]
Then the $U$ and $g$ defined above satisfy the minimization constraints, which proves  ``$\leq$”.

Next we show the other direction of ``$\geq$". 
Let $U$ and $g$ satisfy $U-X-Y$, and $\mathbb{E}[d(f(X,Y),g(U,Y))] \leq D$.
We try to find a $\tilde{W}$ satisfying the constraints in \Cref{Thm1}, i.e., $\tilde{W}= (W,\hat{Z}_{\mathcal{Y}})$ such that $\tilde{W}-X-Y$, $X \in W \in \mathcal{P}(\mathcal{X})$, and $\mathbb{E}[d(f(X,Y),\hat{Z}_Y)] \leq D$.
For any $u\in\mathcal{U}$, define a set 
\begin{equation*}
w(u) = \{ x \in \mathcal{X}: p(u,x) > 0 \},  \label{proof-def-w(u)}
\end{equation*}
and let $w=w(u)$, which implies $X \in W \in \mathcal{P}(\mathcal{X})$. 
Let $\tilde{w}=(w,\hat{z}_{\mathcal{Y}})$. For $(u,x,y)$ such that $p(u,x,y)>0$, define the conditional probability by 
\begin{align}
    p(\tilde{w}|u,x,y)= \mathds{1}\{w = w(u)\} \cdot \prod_{y' \in \mathcal{Y}}\mathds{1}\{\hat{z}_{y'} = g(u,y') \}. \label{proof-def-cprob}
\end{align}
We see that $p(\tilde{w}|u,x,y)$ is a function of $u$ and independent of $(x,y)$, which implies the Markov chain $\tilde{W}-U-X-Y$.
Then we have
$I(U;X|Y)\geq I(\tilde{W};X|Y)$ by the data processing inequality. 
Now it remains to show the distortion constraint. 
In light of \eqref{proof-def-cprob},
we have 
\begin{align*}
    &\hspace{-0.2cm}\mathbb{E}[d(f(X,Y),\hat{Z}_Y)] \\
    &= \sum_{u,x,y}p(u,x,y) \sum_{w,\hat{z}_{\mathcal{Y}}}p(w,\hat{z}_{\mathcal{Y}}|u,x,y)d(f(x,y),\hat{z}_y) \\
    &= \sum_{u,x,y}p(u,x,y)  d(f(x,y),g(u,y)) \\
    &= \mathbb{E}[d(f(X,Y),g(U,Y))] \\
    &\leq D,
\end{align*}
where the last inequality follows by assumptions on $U$ and~$g$. This completes the proof. 

 
In the second part of the proof, if we further suppose the cardinality bound in~\eqref{cardinality bound}, then we have by \eqref{proof-def-cprob} that 
\[
\tilde{W} = (w(U),(g(U,y'))_{y' \in \mathcal{Y}})
\]
is a function of $U$ and each $u$ satisfying $p(u)>0$ is mapped to at most one $\tilde{w}$ such that $p(\tilde{w})>0$. Then the support of $\tilde{W}$ satisfies 
\begin{equation}\label{Thm1cardinality}
    |\{\tilde{w}|p(\tilde{w})>0 \}| 
    \leq |\{u|p(u)>0\}|
    \leq |\mathcal{U}| 
    \leq |\mathcal{X}|+1.
\end{equation}
\end{IEEEproof}

~

\begin{IEEEproof}[Proof of \Cref{Thm2}]
By \Cref{Thm1}, we need to show
\[
\min I(X;\tilde{W}_1|Y) = 
\min I(X;\tilde{W}_2|Y),
\]
where $\tilde{W}_1$ and $\tilde{W}_2$ are defined in \Cref{Thm1,Thm2}, respectively.

We first prove the "$\leq$" direction.
Suppose $\tilde{W}_2 = \hat{Z}_{\mathcal{Y}}$
satisfies $\tilde{W}_2-X-Y$ and $\mathbb{E}[d(f(X,Y),\hat{Z}_Y)] \leq D$.
For any $\tilde{w}_2 = \hat{z}_{\mathcal{Y}} \in \hat{\mathcal{Z}}^{\mathcal{Y}}$, let
\begin{equation}
    w(\tilde{w_2}) = \{ x \in \mathcal{X}: p(\tilde{w}_2,x) > 0 \}. \label{proof-def-w-w2}
\end{equation}
We then define 
\begin{equation}
\tilde{W}_1 = (w(\tilde{W}_2),\tilde{W}_2). \label{proof-def-w1-tilde}
\end{equation}
Since  $\tilde{W}_1$ is a function of $\tilde{W}_2$, we have the Markov chain $\tilde{W}_1-\tilde{W}_2-X-Y$ which by the data processing inequality implies that $I(\tilde{W}_1;X|Y) \leq I(\tilde{W}_2;X|Y)$. 
Moreover, we have  $\mathbb{E}[d(f(X,Y),\hat{Z}_Y)] \leq D$ by the assumption on $\tilde{W}_2$.

For any $\tilde{w}_1$ satisfying $p(\tilde{w}_1)>0$, there exists a $\tilde{w}_2$ such that $\tilde{w}_1 = (w(\tilde{w}_2),\tilde{w}_2)$ and $p(\tilde{w}_2)>0$. 
In addition, there exists an $x$ such that $p(\tilde{w}_2,x)>0$, from which we see that $w(\tilde{w}_2)$ is not empty by its definition in \eqref{proof-def-w-w2}.
By setting $w = w(\tilde{W}_2)$, we have $W \neq \emptyset$ with probability $1$ and then $W \in \mathcal{P}(\mathcal{X})$.

For $(w,x)$ satisfying $p(w,x)>0$, there exists a $\tilde{w}_1 = (w,\hat{z}_{\mathcal{Y}})$ such that $p(\tilde{w}_1,x)>0$, 
which implies the existence of $\tilde{w}_2$ such that $p(\tilde{w}_1, \tilde{w}_2,x)>0$. 
By the definition of $\tilde{W}_1$ in~\eqref{proof-def-w1-tilde}, we have $\tilde{w}_1 = (w(\tilde{w}_2),\tilde{w}_2)$. Then by $p(\tilde{w}_2,x) > 0$, and the definition in \eqref{proof-def-w-w2}, we have $x \in w(\tilde{w}_2) = w$, which proves that $X \in W$ and thus the ``$\leq$" direction.


Next we prove the "$\geq$" direction. Let $\tilde{W}_1 = (W,\hat{Z}_{\mathcal{Y}})$
such that $\tilde{W}_1-X-Y$, $X \in W \in \mathcal{P}(\mathcal{X})$, and $\mathbb{E}[d(f(X,Y),\hat{Z}_Y)] \leq D$.
We directly define $\tilde{W}_2 = \hat{Z}_{\mathcal{Y}}$ which is a function of $\tilde{W}_1$ and satisfies $\tilde{W}_2-\tilde{W_1}-X-Y$, $\mathbb{E}[d(f(X,Y),\hat{Z}_Y)] \leq D$. 
The inequality $I(\tilde{W}_1;X|Y)\geq I(\tilde{W}_2;X|Y)$ is obtained by the data processing inequality.
This completes the proof.
\end{IEEEproof}

~

\begin{IEEEproof}[Proof of \Cref{Thm3}]
Applying \Cref{Thm1} to the case that $D = 0$, we only need to prove 
$$\min_{\substack{W-X-Y \\ X \in W \in \Gamma_d}}I(X;W|Y) = \min I(X;\tilde{W}|Y),$$
where the minimum on the right hand side is taken over all random variables $\tilde{W} = (W',\hat{Z}_{\mathcal{Y}})$,
satisfying $\tilde{W}-X-Y$, $X \in W' \in \mathcal{P}(\mathcal{X})$, 
and 
\begin{align}
    d(f(x,y),\hat{z}_y) \!= 0 \text{ for }\tilde{w}\! = (w',\hat{z}_{\mathcal{Y}}) \text{ s.t. }p(\tilde{w},x,y)>0. \label{proof3-w-tilde-condition}
\end{align}

First we prove the "$\leq$" direction.
For any $\tilde{W} = (W',\hat{Z}_{\mathcal{Y}})$ satisfying the constraints of the right hand side, let 
\begin{equation}
w(\tilde{w}) = \{x \in \mathcal{X}:p(\tilde{w},x)>0\}.
\label{proof-def-w-tildew}
\end{equation}
Then we define  
\begin{equation}
w = w(\tilde{W}),
\label{proof-def-w}
\end{equation}
which implies the Markov chain $W-\tilde{W}-X-Y$ and thus we have $I(W;X|Y) \leq I(\tilde{W};X|Y)$ by the data processing inequality.

For any $w$ such that $p(w)>0$, there exists a 
\begin{equation}
    \tilde{w} = (w',\hat{z}_{\mathcal{Y}})  \label{proof3-def-w-tilde}
\end{equation}
such that $p(\tilde{w})>0$ and $w = w(\tilde{w})$ by \eqref{proof-def-w}. So for any $y \in \mathcal{Y}$ and  $x \in w = w(\tilde{w})$ such that $p(x,y)>0$, we have $p(\tilde{w},x) >0$ by the definition in~\eqref{proof-def-w-tildew}.
Since $\tilde{W}-X-Y$ is a  Markov chain, we have 
\[
p(\tilde{w},x,y) = \frac{p(\tilde{w},x)p(x,y)}{p(x)}>0,
\]
which together with \eqref{proof3-w-tilde-condition} and \eqref{proof3-def-w-tilde} implies $d(f(x,y),\hat{z}_y) = 0$. So we have $w_y \subseteq B(\hat{z}_y,0)$ (c.f. Definition~2) for any $y \in \mathcal{Y}$. Then by the assumption that $p(w)>0$, we have $W \in \Gamma_d$. 

Moreover, for any $x$ such that $p(w,x)>0$, there exists a $\tilde{w}$ such that $p(w,\tilde{w},x)>0$, which implies $p(\tilde{w},x)>0$. 
Then by \eqref{proof-def-w-tildew} and \eqref{proof-def-w}, we have $X \in W$. 
This proves the ``$\leq$" direction.


Next we show the "$\geq$" direction.
Let $W$ be a random variable satisfying $W-X-Y$ and $X \in W \in \Gamma_d$. 
For any $w \in \Gamma_d$ and $y' \in \mathcal{Y}$, there exists a $\hat{z}_{w,y'}$ such that 
$w_{y'} \subseteq B(\hat{z}_{w,y'},0)$, i.e., $d(f(x,y),\hat{z}_{w,y'}) = 0$ for any $x \in w$ satisfying $p(x,y')>0$. 

We then define the conditional probability of $\tilde{W} = (W',\hat{z}_{\mathcal{Y}})$ given $(W,X,Y)$ to be
\[
p(w',\hat{z}_{\mathcal{Y}}|w,x,y) = \mathds{1}\{w' = w\} \cdot \prod_{y' \in \mathcal{Y}} \mathds{1}\{\hat{z}_{y'} = \hat{z}_{w,y'} \}.
\]
Then the Markov chain $\tilde{W}-W-X-Y$  holds, which by the data processing inequality implies that $I(X;W|Y)\geq I(X,\tilde{W}|Y)$. 
Furthermore, it is easy to see that $X \in W' \in \mathcal{P}(\mathcal{X})$ from $X \in W \in \Gamma_d$.
Finally, for $\tilde{w} = (w',\hat{z}_{\mathcal{Y}})$ such that $p(\tilde{w},x,y')>0$, we have $x \in w' \in \Gamma_d $ and $\hat{z}_{y'} = \hat{z}_{w',y'}$, which imply that 
\[
d(f(x,y'),\hat{z}_{y'}) = d(f(x,y'),\hat{z}_{w',y'}) = 0.
\]
This proves the ``$\geq$" direction and thus the theorem.
\end{IEEEproof}

%% file: A5.tex
\section{Conclusion}\label{sec:conclusion}
In this paper, we proposed a characteristic multi-hypergraph for the lossy computing problem with side information. 
The graph-based rate-distortion function was characterized. 
Two reductions of the multi-hypergraph for reducing the support of the auxiliary random variable and for the zero distortion case were studied. 
We also demonstrated through an example how to  design graph-based coding schemes and compute the rate-distortion function in light of the multi-hypergraph.
Possible generalizations of the multi-hypergraph to other multi-terminal computing problems are under investigation. 